\documentclass{gPHT2e}
\usepackage{xcolor}

\usepackage{endnotes}
\let\footnote=\endnote

\begin{document}

\jvol{00} \jnum{00} \jyear{2012} \jmonth{December}

\title{{\itshape Relaxors, spin-, Stoner- and cluster-glasses}}

\author{David Sherrington
$^{\ast}$\thanks{$^\ast$ Email: d.sherrington1@physics.ox.ac.uk
\vspace{6pt}}
\\\vspace{6pt}
{\em{ Rudolf Peierls Centre for Theoretical Physics, 1 Keble Rd., Oxford OX1 3NP, UK;}}
\newline
{\em{ Santa Fe Institute, 1304 Hyde Park Rd., Santa Fe, NM 87501, USA
\newline
}}
}

\maketitle

\begin{abstract}
It is argued that  the main characteristic features of  displacive relaxor ferrolectrics of
the form ${\rm{A(B,B')}\rm{O}}_3$ with isovalent ${\rm{B,B'}}$ can be explained and understood in terms of a soft-pseudospin analogue of conventional spin glasses as extended to itinerant magnet systems. The emphasis is on conceptual comprehension and on  stimulating new  perspectives with respect to previous and future studies. Some suggestions are made for further studies both on actual real systems  and on test model systems to probe further.
The case of heterovalent systems is also considered briefly.

\medskip

\medskip

\end{abstract}

\section{Introduction}

There has been much interest in relaxor ferroelectrics since their discovery in the 1950s by the group of George Smolenskii in the USSR \cite{Smolenskii}, following interest in pure displacive ferroelectrics of perovskite structure and make-up ${AB\rm{O}}_3$, where $A$ is an ion of nominal charge +2, $B$ an ion of charge +4 and O is oxygen, of charge -2. In particular, a new class of alloy materials was discovered in $\mathrm {Pb(Mg}_{1/3}\mathrm{Nb}_{2/3}\mathrm{)O}_3$ (PMN)
 with interesting
 frequency-dependent peaks in its dielectric susceptibilities as a function of temperature \cite{Smolenskii2}, just below room temperature, but without any macroscopic polarization in the absence of applied fields.  It has further turned out that  there are 
 manifestations of 
nano-scale polar domains 
{\textcolor{black}{(polar nano-regions, PNRs)}} \cite{Egami}, beneath a higher `Burns' temperature  \cite{Burns} at which deviations from high-temperature paraelectricity become apparent, 
{\textcolor{black}{and}
 non-ergodicity {\textcolor{black}{onsets}   beneath a temperature comparable with that of the
susceptibility peaks \cite{Kleemann}, \cite{Kleemann_FC_ZFC}, \cite{Levstik}. However, more than half a century later than the discovery of these systems, there remains a lack of  full understanding and agreement on the microscopic physical origin of the observed effects.

Recently, the present author proposed  \cite{Sherrington2013, SherringtonPSSB} that a conceptual explanation
 may be found     in analogy with a soft-spin variant of a conventional spin glass \cite{Binder, Mydosh, Fischer}, with the 
{\textcolor{black}{glassy onset of non-ergodicity without global lattice change the analogue of a spin glass transition and the}}
PNRs manifestations of localized state precursors of macroscopic order \cite{Sherrington-Mihill, Sherrington-Mihill2}, especially simply for a more recently discovered class of isovalent relaxors, such as $\mathrm {Ba(Zr}_{1-x}\mathrm{Ti}_{x})\mathrm{O}_3$ (BZT) in which the B-site ions are occupied randomly by Ti or Zr ions, both of charge +4. BZT is conceptually clearer than heterovalent PMN in which the +4 Ti-ions of $\mathrm {PbTi}\mathrm{O}_3$ are replaced randomly by a mixture of one third Mg +2 and two thirds Nb +5, giving rise also to \textcolor{black}{significant} effective random fields {\textcolor{black}{relative to the  charge distribution of the underlying ${AB\rm{O}}_3$ perovskite template}}  \footnote{\textcolor{black}{Note that in this paper  the expression `field' is reserved for the coefficients of terms of linear order in the Hamiltonian measuring the energy cost of displacements of ions away from their pure perovskite template. Some other authors effectively  consider  minimization with respect to part of the Hamiltonian and use the resulting polarization as providing effective fields on the rest of the Hamiltonian.}}.

The author's paper on BZT \cite{Sherrington2013} was based on the premise that the key features of the relaxor behaviour of BZT can be understood by analogy with  experimental spin glasses, as a compromise between frustrated intersite interactions and quenched site-to-site variations in the extent to which local displacements have propensities in the absence of interactions towards retaining their high-temperature perovskite structure or moving away from these positions. 
{\textcolor{black}{In addition, the strength of the binding needs to be sufficient to overcome the local displacive energy cost.}}
 More specifically it was argued that for displacive relaxors the glassy ordering is analogous to
{\textcolor{black}{earlier proposals for}} 
itinerant glassy magnets such as {\bf{Rh}}Co \cite{Coles}, essentially 
 a combination of the Stoner-glass concept introduced in \cite{Hertz1979} and  a cluster glass with the PNRs analogues of the statistically cluster-localized moments proposed in \cite{Sherrington-Mihill, Sherrington-Mihill2}.

\cite{Sherrington2013} was stimulated in part by a prior computer simulation study of BZT \cite{Akbarzadeh2012} for the 50/50 concentration case, with observations in that computer study explainable by   theoretical modelling and analogy.
The simulational studies observed both PNRs and ergodicity-breaking. The prediction in \cite{Sherrington2013}  that PNRs above the relaxor onset temperature (the low-frequency susceptibility peak temperature) but beneath a higher temperature can be viewed as statistically clustered localized phonons was given posterior support in experimental observations  of \cite{Manley}.

However, not all 
researchers in the topic
share the author's vision  and 
{\textcolor{black}{prefer}} 
 other pictures, such as random fields  as the driving mechanism  \cite{Kleemann, Kleemann2014a}, even for the isovalent relaxors \cite{Kleemann2014b}\footnote{\textcolor{black}{See also the second sentence of the previous note.}}. Also,  the non-observation, at least to date, of relaxor behaviour in $\mathrm {Pb(Zr}_{1-x}\mathrm{Ti}_{x})\mathrm{O}_3$ has been suggested as a counter-example to the spin-glass analogy 
{\textcolor{black}{for an isovalent alloy} \cite{Gehring1, Gehring2}.

Hence, in this communication the author attempts  to explain further his reasoning, indicating also some of the similarities and overlaps of his modelling with other descriptions, as well as some of the the differences, in the hope of clarifying the potential conceptual issues. The mathematical models discussed below are similar to those used by previous authors concerned with first-principles quantum-mechanical evaluation of effective classical Hamiltonians followed by computer simulation of their statistical mechanics \cite{Zhong, Akbarzadeh2012,Prosandeev}. However, the approach used below is {\textcolor{black}{complementary;} minimalist and qualitative, using simplified Hamiltonians, approximations and analogies, with the aim of  probing possibilities for understanding and suggesting opportunities for the incorporation of unconventional concepts in future thinking, experimentation and simulation, rather than concerned with precision or quantification.

\section{Microscopic `moments'}
In ferroelectrics (and antiferroelectrics and relaxors) systems are often classified as `order-disorder' or as `displacive', depending upon whether the macroscopic ordering is of discrete or continuous variables. For example, $\mathrm{NaN0}_{2}$ is considered as an order-disorder system since well-defined (discrete) local dipolar moments exist already above the ferroelectric ordering temperature, while
$\mathrm {Ba}\mathrm{Ti}\mathrm{O}_3$ is usually described as a displacive ferroelectric since, to a first approximation, there is no local 
\textcolor{black}{electric}
moment above the ferroelectric transition temperature but rather it grows continuously {\textcolor{black}{from zero}} as the temperature is reduced {\textcolor{black}{through the transition}\footnote{In fact there is evidence of a smaller order-disorder component as well as the main displacive one \cite{Zhong} but the discussion below allows straightforwardly for such a scenario.}.

Correspondingly for magnetic systems one has (i) order-disorder transitions  where the local magnetism is due to well-defined local moments, typically associated with incomplete d- and f-shell atoms or ions, (ii) an analogue of displacive transitions  in (metallic) itinerant magnets where the magnetism is due to Coulombic interactions in conduction-electron bands, typically in d-electron metals, displacing the symmetry of up and down conduction bands. 

Although order-disorder and displacive transitions are often viewed as separate entities, in fact they can be considered as continuously related through variation of suitable parameters. Thus, if the non-interacting local displacement contribution to a Hamiltonian takes the form
\begin{equation}
H_L = \kappa u^2 +\lambda u^4
\label{local soft spin}
\end{equation}
with $\lambda$ positive, then its minimum will be at $u=0$ for $\kappa$ positive but for $\kappa$ negative it has two minima at $u=\pm (-\kappa/2\lambda)^{1/2}$ \footnote{For simplicity Equation (\ref{local soft spin}) is restricted to scalar variables. Already in earlier numerical/simulational studies, such as \cite{Zhong} and \cite{Akbarzadeh2012}, a more complete form is considered, but recall that the aim here is illustrative comparison for conceptualization, not numerical accuracy {\it{per se}}.}. When site-to-site interaction effects are included, permitting a cooperative ordering, 
$\kappa$ positive
corresponds to a `displacive' transition, whereas 
$\kappa$ negative
 corresponds to `order-disorder'.
 The order-disorder case sharpens as the limits ($\kappa \to -\infty$, $\lambda \to \infty$, with $(\kappa/\lambda) \to $ const) are taken
\footnote{{\textcolor{black}{Field-theoretic studies of even nominally discrete/hard spin systems are often formulated in terms of a soft-spin formulation such as Equation (\ref{local soft spin}), so as to allow for continuous variables, but finally the limit  ($\kappa \to -\infty$, $\lambda \to \infty$, with $(\kappa/\lambda) \to $ const) is taken.}}}. 
Clearly, as $\kappa$ is taken through 0 one passes continuously between `order-disorder' and `displacive'. This explains why experimentally some systems undergo phase transitions with a mixture of order-disorder and displacive aspects, as corresponding to small negative {$\kappa$}.

As an example of the `displacive' modelling of an itinerant magnet, pure or alloy, one can consider the procedure described in \cite{Sherrington-Mihill, Sherrington-Mihill2}. With some simplifying assumptions \footnote{Retaining spin fluctuations but ignoring charge fluctuations and assuming uniformity of the resultant effective bare band structure.}    one may start from the Hubbard model
\begin{equation}
H=\sum_{ij,\sigma}t_{ij}a_{i\sigma}^\dagger a_{j\sigma} -{\frac{1}{4}}\sum_{i}U_{i}{\textit{\textbf{S}}}_{i}.{\textit{\textbf{S}}}_{i},
\end{equation}
where the $a,a^{\dagger}$ are site- and spin-labelled d-electron annihilation and creation operators,
${\bf{S}}_{i} =
\sum_{s,s' = \pm 1}  a^{\dagger}_{is}{\bm{\sigma}}_{ss'}a_{is'} $,
where ${\bm{\sigma}}_{ss'}$ is the Pauli matrix, and
allowance is made for different atom types with different Coulomb repulsions $U$. For simplicity below, discussion is restricted to just two types of atom, one with $U=0$ and one with a finite $U$. Expressing the partition function in terms of a Grassmann functional integral in  $a,a^{*}$ fields, linearising in terms of single sites by utilising  the inverse of completing the square \cite{Muhlschlegel-Zittartz63, Sherrington67}
\begin{equation}
\exp(a^{2})={\pi^{-1/2}}\int dx \exp(-x^{2}-2ax),
\label{Completing_square}
\end{equation}
via an effective local magnetization field,
perturbation-expanding in the magnetization field by  integrating out the electron variables and
taking the static approximation, one obtains an effective magnetization Hamiltonian
\begin{eqnarray}
H_{M}= \sum_{i}  (U^{-1}-\chi_{ii})
|{\textit{\textbf{M}}}_{i}|^2 -
 \sum_{ij;i\neq j}\chi_{ij}{\textit{\textbf{M}}}_{i}.{\textit{\textbf{M}}}_{j} \nonumber \\
 -
 \sum_{ijkl;\alpha\beta\gamma\delta}\Pi^{\alpha\beta\gamma\delta}_{ijkl}
M^{\alpha}_{i}M^{\beta}_{j}M^{\gamma}_{k}M^{\delta}_{l} + ...
\label{H_SM2}
\end{eqnarray}
where ${\bf{M}}_{i}=U_{i}{\bf{m}}_{i} = U_{i} <a^{\dagger}_{is}{{{\boldsymbol{\sigma}}}}_{ss'}a_{is'}>$, the $\chi_{ij}$ and $\Pi_{ijkl}$ are the two-site susceptibilities and analogous four-site
 functions of the bare conduction-band structure, and the sums are over only sites with $U_{i} \neq 0$. The local terms in this expression are
\begin{eqnarray}
H_{ML}= \sum_{i} (U^{-1}-\chi_{ii})
|{\textit{\textbf{M}}}_{i}|^2
 -
 \sum_{i;\alpha\beta\gamma\delta}\Pi^{\alpha\beta\gamma\delta}_{iiii}
M^{\alpha}_{i}M^{\beta}_{i}M^{\gamma}_{i}M^{\delta}_{i} + ...,
\label{H_SM3}
\end{eqnarray}
again over only sites with $U \neq 0$,
which is immediately recognisable as being of the form of Equation (\ref{local soft spin}), with
$\kappa \sim  (U^{-1}-\chi_{ii})$ and $\lambda \sim -\Pi_{iiii} {\textcolor{black}{>0}}$. It  immediately yields the standard Anderson mean-field criterion ($(1-U\chi_{ii}) <0$)  for a local moment on an isolated ion with $U \neq 0$ in a sea of ions with $U=0$ \cite{Anderson61}.

\section{Interactions}

To exhibit phase transitions a system needs interactions as well as the local terms. For the case of the  Hubbard model example for itinerant magnets these have already been exhibited in the extra terms in Equation (\ref{H_SM2}) compared with Equation (\ref{H_SM3}). The simplest interaction term is
\begin{equation}
H_{MI}= \sum_{ij;i\neq j}\chi_{ij}{\textit{\textbf{M}}}_{i}.{\textit{\textbf{M}}}_{j}.
\label{M interaction}
\end{equation}
This gives the condition for itinerant ferromagnetism in a simple pure system (all sites with the same $U$) as
\begin{equation}
1-U\sum_{J}\chi_{ij} < 0,
\label{M interaction}
\end{equation}
the well-known Stoner criterion.

For displacive ferroelectrics the microscopic origin of terms like Equation (\ref{M interaction}) is different but the consequence is similar.
Interactions occur both because of shortish-range quantum mechanical reasons, estimable by first principles density functional methods,
and because of longer-range Coulombic interactions  \cite {KingSmith}. Restricting to two-site interactions and ignoring coupling to overall strains, these interaction terms may be subsumed to give the Hamiltonian schematically as \footnote{For an explicit discussion  see \cite{Zhong}.}
\begin{equation}
\begin{split}
H_{DF}=\sum_{i}
\{\kappa_{i}| \textit{\textbf{u}}_{i} |^2
+\lambda_{i}|\textit{\textbf{u}}_{i}|^4
+\gamma_{i} [u_{ix}^{2}u_{iy}^{2} +u_{iy}^{2}u_{iz}^{2} +u_{iz}^{2}u_{ix}^{2}]
\}\\
+ {{1/2}\sum_{(i,j;i \neq j)} H_{int}({\textit{\textbf{u}}}_{i}, {\textit{\textbf{u}}}_{j},  {\textit{{\textbf{R}}}}_{ij})}.
\label{eq:HDF}
\end{split}
\end{equation}

To attain cooperative order in a system with positive {$\kappa$} there needs to be a sufficient lowering of the free energy due to the interaction to overcome the local displacement penalty. This is the direct analogue of the condition for magnetic order given in the itinerant  magnetic example above.

Thus, viewed this way, there is conceptual similarity between itinerant magnets and displacive ferroelectrics, a similarity that extends to more complex order, such as antiferroelectricity and charge-density-wave magnetism, if the the form of the interaction is such that (free) energy is lowest for such ordering. Both require a minimum interaction
binding energy gain relative to the harmonic displacement energy cost in order to bootstrap cooperative order. In both itinerant magnetic and displacive electric cases there are well-known examples of pure systems 
{\textcolor{black}{which satisfy  this requirement}} 
 and others 
{\textcolor{black}{which do not;}} 
thus $\mathrm {Ba}\mathrm{Ti}\mathrm{O}_3$ is a ferroelectric at low enough temperature while $\mathrm {Ba}\mathrm{Zr}\mathrm{O}_3$ remains paraelectric at all temperatures (presumably because its $\kappa$ is too large), and analagously Co is a ferromagnet while pure Pd is only paramagnetic.

In displacive systems there are also couplings to the overall strain. These are relevant for systems with periodic ordering, such as ferroelectricity, permitting also changes in $c/a$ ratios, but are not so important in relaxor systems where the overall symmetry is not affected. Hence they are ignored in the main discussion here.

\section{Alloys}

The interest  of this article is principally in alloys which exhibit glassiness, a broad range of timescales extending to very slow. This is evident in relaxor systems in the fact that the peaks in the susceptibility are significantly frequency-dependent, their temperatures increasing with frequency. This is in contrast to conventional transitions, such as that to ferroelectricity, where  the susceptibility peak positions are little affected by the frequency of the probe.

Small amounts of alloying, randomly replacing  host atoms by other elements, retain the main characteristics of the pure system transition, such as the independence of the susceptibility peak on frequency. Larger amounts of admixture in systems with frustration can (and, it is argued here, often do)  lead to a relaxor (glass)  state.

As was pointed out in \cite{Sherrington2013}, in analogy with well-known behaviour in conventional (local moment) spin glasses, the combination of frustrated interactions and locally variable $\kappa$ provide ingredients known to lead to spin glass behaviour, particularly when one of the constituents is essentially immobile, having a $\kappa$ that is too large (positive) to permit periodic ordering as a pure material.

Canonical experimental local moment spin glasses, such as {\bf{Au}}Fe, in which one ingredient is magnetic (here Fe) and the other non-magnetic (here Au) have Hamiltonians of the form
\begin{equation}
H_{SG}=-{1/2}\sum_{(i,j; i \neq j)(Mag)}  {J}({\textit{\textbf{R}}}_{ij}){\textit{\textbf{S}}}_{i}.{\textit{\textbf{S}}}_{j}
\label{eq:Hcsg}
\end{equation}
where the ${\it\bf{S}}_{i}$ are localised spins on the magnetic atoms and the summation is restricted to the corresponding sites. The exchange interaction ${J}({\textit{\textbf{R}}}_{ij})$ is frustrated; in the case of a simple metallic system like {\bf{Au}}Fe as the RKKY form $J{\cos(2k_{F}(R))}
/R^{3}$, which provides spatial frustration through its oscillations in sign; while in insulating spin glasses {\textcolor{black}{in some cases}} the range of interaction is shorter but still spatially frustrated through antiferromagnetic terms beyond nearest neighbour, {\textcolor{black}{as, for example, in ${\textrm{Eu}}_{x}{\textrm{Sr}}_{(1-x)}{\textrm{S}}$, and in others there is frustration of  long-range dipolar character, for example in 
${\textrm{LiHo}}_{x}{\textrm{Y}}_{(1-x)}{\textrm{F}}_{4}$}
\footnote {In fact purely antiferromagnetic interactions can be frustrated if of longer than nearest neighbour range.}.

Let us next consider
displacive alloys ${A(B',B')}O_3$ in which the $B'$, $B''$ are isovalent (both 4+) and, for concise simplicity, assume that the effects of the $A$ and $O$ ions are absorbed into an effective interaction between the $B$ ions. Each of the $B$-types has its own value of $\kappa$. If one of them (say that of $B'$) is sufficiently large such that when it is the only $B$-constituent the material is paraelectric at all temperatures, \textcolor{black}{then its displacements can be ignored}
\textcolor{black}{and} the Hamiltonian can be written as a sum over only the other $B$-sites ($B''$) 
\begin{equation}
H_{DR}=\sum_{i}
\{\kappa_{i}| \textit{\textbf{u}}_{i} |^2
+\lambda_{i}|\textit{\textbf{u}}|^4
\}+ {{1/2}\sum_{i,j; i \neq j} H_{int}^{B eff}({\textit{\textbf{u}}}_{i}, {\textit{\textbf{u}}}_{j},  {\textit{\textbf{R}}}_{ij})}
\label{eq:HDR}
\end{equation}
where the summations are now over $B''$-sites only
\footnote{The local fourth-order cross-terms {\textcolor{black}{and higher-order local terms}}  are not written explicitly. 
.}. 
{\textcolor{black}{Global strain coupling is ignored here since it has been observed that the average cubic perovskite structure  is maintained in  transitions from paraelectric to relaxor}}
\footnote{
In their simulation of BZT(50/50) Akbarzadeh et al. \cite{Akbarzadeh2012}
demonstrated explicitly that the strain terms are relatively less important in the relaxor phase of BZT.}.
In the case of BZT the summations are thus only over Ti sites. The interaction term
$H_{int}^{B eff}({\textit{\textbf{u}}}_{i}, {\textit{\textbf{u}}}_{j},  {\textit{\textbf{R}}}_{ij)}$ is spatially frustrated but for the pure Ti case the minimum of $H_{DR}$ is ferroelectric.

{\textcolor{black}{To second order in the displacements the interaction term may be expressed in the form 
\begin{equation}
H_{int}({\textit{\textbf{u}}}_{i}, {\textit{\textbf{u}}}_{j},  {\textit{\textbf{R}}}_{ij}) =
-{1/2}\sum_{(i,j; i \neq j; \alpha, \beta )}J_{\alpha \beta}({\textbf{R}}_{ij}) u_{i,\alpha}u_{j, \beta}
\end{equation}
where $J_{\alpha \beta}({\textbf{R}}_{ij})$ is made up of  terms of shortish range but extending beyond nearest neighbour, calculable by first-principles quantum analysis, and others of long-ranged dipolar character \cite{Zhong}.  Both involve frustration with terms of opposing character.
Given the evidence from magnetic spin glasses that many types of frustration  can lead to a spin glass state  in the presence of sufficient quenched disorder, it seems reasonable to expect that a similar situation can occur in  displacive ferroelectrics if their fundamental interactions are frustrated, there is sufficient quenched disorder such that the potential binding energy gain is greatest for a  distorted state without periodic order, and the potential binding energy of spontaneous distorsion is sufficient to overcome the local displacement energy costs.   This expectation for BZT is in accord the simulations of Akbarzadeh {\it{et. al}} \cite{Akbarzadeh2012} on BZT 50/50, showing the characteristic spin glass signature of   a separation between field cooled (FC) and zero-field cooled (ZFC) quasi-static susceptibilities beneath a critical temperature, and the experiments of several groups \cite{Maiti, Kleemann2009} showing significant frequency-dependence in peaking of the finite-frequency susceptibilities.}}

{\textcolor{black}{In comparison one might note that for the itinerant alloy system also, the interaction term $\chi_{ij}$ is spatially frustrated, in the same Fermi-surface-driven  manner as the RKKY interaction.}

\section{ Mean field theory, free energy, polar nanoregions, localization and phase transitions}
\label{Localization}

In this section we consider mean field solutions for displacive alloys. Because of the spatial disorder in Equation (\ref{eq:HDR}) mean field theory must allow for spatial variations in $<{\bf{u_{i}}}>$. For zero temperature one can minimize $H_{DR}$, but for finite temperature one should minimise the corresponding free energy, which we shall write as
\begin{equation}
F_{DR}=\sum_{i}
\{\tilde{\kappa}_{i}(T)| \textit{\textbf{u}}_{i} |^2
+\tilde{\lambda}_{i}(T)|\textit{\textbf{u}}|^4
\}+ {\sum_{ij...}' F_{int}({\textit{\textbf{u}}}_{i}, {\textit{\textbf{u}}}_{j},...  {\textit{\textbf{R}}}_{ij...})}
\label{eq:FDR}
\end{equation}
and, for simplicity, further concentrate on bi-linear contributions to $F_{int}$ and scalar $u$; \footnote{These simplifications are not crucial to the conceptual argument.}
\begin{equation}
F_{SDR}=\sum_{i}
\{\tilde{\kappa}_{i}(T) u_{i} ^2
+\tilde{\lambda}_{i}(T)u_{i}^4
\}
- {{1/2}\sum_{i,j; i \neq j}\tilde{J}(R_{ij}, T)u_{i}u_{j}}.
\label{eq:SFDR}
\end{equation}
All the coefficients are temperature-dependent, but the most important for our purpose is  $\tilde{\kappa}_{i}$ which is expected to increase with temperature, making bootstrapping order more difficult as temperature is increased and yielding paraelectricity as the high temperature phase. {\textcolor{black}{It is assumed that}} $\tilde{\lambda}$ remains positive throughout \footnote{\textcolor{black}{If this is not the case then terms of higher order must be included, {\it{e.g.} $\gamma u^{6}$}}}. Again for simplicity, the temperature dependence of $\tilde{\lambda}$ and $\tilde{J}$ (or $F_{int}$) will be suppressed in the discussion.

Minimizing $F_{SDR}$, Equation (\ref{eq:FDR}),  with respect to the $\{u_{i}\}$ yields
\begin{equation}
-2\tilde{\kappa}_{i}(T) u_{i}
+ {\sum_{j}\tilde{J}(R_{ij})u_{j}} = 4\tilde{\lambda}_{i}u_{i}^3  .
\label{eq:Min F(u)}
\end{equation}
Clearly, this equation always allows solutions $\{{u=0}\}$, corresponding to undisplaced paraelectricity, but interest is in possible solutions $\{{u \neq 0}\}$. Such {\textcolor{black}{$\{{u \neq 0}\}$}} solutions only occur for small enough {\textcolor{black}{$\{\tilde{\kappa}\}$ such that the $\tilde{J}$ term can bootstrap-yield a right hand side (r.h.s.) of Equation (\ref{eq:Min F(u)}) that has the same sign as $u_{i}$}}. {\textcolor{black}{They}} can be either {\textcolor{black}{extended or localized}},
{\textcolor{black}{depending upon whether $\{u \neq 0\}$ sites span the whole system or not}}. {\textcolor{black}{Localized states correspond to PNRs while a true phase transition requires a globally extended solution.}}

{\textcolor{black}{A potentially helpful}} 
way to understand these possibilities is to compare Equation (\ref{eq:Min F(u)}) with the  Anderson (eigenfunction) localization  equation \cite{Anderson58}
\begin{equation}
\epsilon_{i} \phi_{i}
+ {\sum_{j}t_{ij}\phi_{i}} = E\phi_{i},
\label{eq:Anderson loc}
\end{equation}
with the identifications
\begin{equation}
\epsilon_{i} =-\tilde{\kappa_{i}}(T);  \hspace{1 cm}
t_{ij}={\tilde{J}}(R_{ij})/2,
\label{eq:Identification}
\end{equation}
{\textcolor{black}{corresponding to diagonalization of the r.h.s. of Equation (\ref{eq:Min F(u)})}}.
This 
{\textcolor{black}{shows}} that that \textcolor{black}{the condition for}} non-zero solutions to Equation (\ref{eq:Min F(u)}) corresponds to \textcolor{black}{that for} $E \ge 0$ solutions to Equation (\ref{eq:Anderson loc}). 
{\textcolor{black}{Thus, if $E < 0$ the system remains paraelectric with all $u_{i}=0$.}}

The density of states $\rho(E)$ of solutions of Equation (\ref{eq:Anderson loc}) of eigenvalue $E$ 
gives insight into the corresponding  
solutions of Equation (\ref{eq:Min F(u)}).
Figure \ref{fig1} shows a schematic density of states for a system such as described by Equation (\ref{eq:Anderson loc}), {\textcolor{black}{for a situation with all eigenvalues negative, corresponding for Equation (\ref{eq:Min F(u)}) {\textcolor{black}{to}} a temperature too high} for $u \neq 0$ solutions, in the paraelectric region. As the temperature is reduced the density of states distribution moves to higher $E$. At a first characteristic temperature the density of states crosses zero and permits non-zero $u$ solutions to Equation (\ref{eq:Min F(u)}).

\begin{figure}
\begin{center}
\resizebox*{10cm}{!}
{\includegraphics{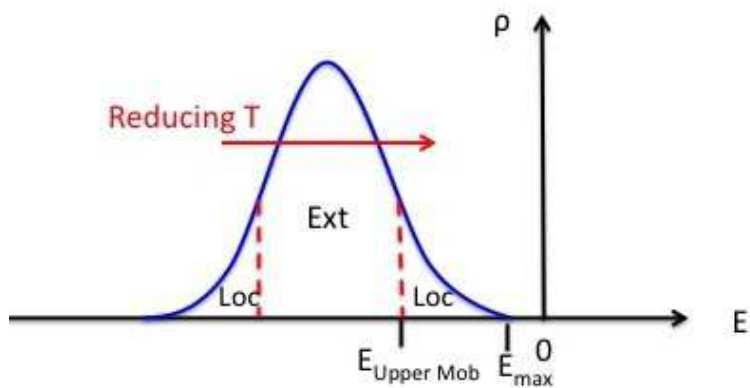}}%
\caption{
\label{fig1}
Schematic density of states of an analogous Anderson model, showing typical localized (Loc) and extended (Ext) states in the case of a disordered system. Also shown is the effect of reducing temperature in the displacive system. Polar nanoregions onset when the upper band edge ($\textrm{E}_{\textrm{max}}$) crosses zero energy and macroscopic order occurs when the  upper mobility edge ($\textrm{E}_{\textrm{{Upper Mob}}}$) crosses zero.}
\end{center}
\end{figure}

For a pure system
the  eigenfunction of Equation (\ref{eq:Anderson loc}) with the largest eigenvalue is extended \footnote{\textcolor{black}{In fact, all the eigenfunctions are extended and the mobility edges, shown schematically in Figure \ref{fig1}, coincide with the band edges.}}.
Thus, with a small enough $\kappa$
there will be a spontaneous phase transition when the highest eigenvalue reaches zero, 
{\textcolor{black}{to a phase with the symmetry of the highest eigenfunction}}. 
{\textcolor{black}{As the temperature is lowered the order parameter of the symmetry breaking will initially grow linearly in ${(E_c)}^{1/2}$ or equivalently${(T_c -T)^{1/2}}$.}}
 $\mathrm {Ba}\mathrm{Ti}\mathrm{O}_3$ (BT) is an example (with the transition to a ferroelectric). 

However, in a disordered system the states near the band edges of Equation (\ref{eq:Anderson loc}) are localized, with extended states only possible within `mobility edges'  (see Figure(\ref{fig1}). Thus, in a disordered system, as the temperature is reduced there first onset metastable localized clusters at a temperature $T_{A}$
at which the upper band edge of the Anderson equation reaches zero energy, with a possible true phase transition only at a lower temperature $T_{c}$ at which {\textcolor{black}{an extended state becomes a solution to Equation (\ref{eq:Min F(u)}), analogous to where}} the upper mobility edge {\textcolor{black}{of the Anderson model}} crosses zero.

In conventional Anderson localization modelling the $t_{ij}$ are unfrustrated and usually short-ranged and the extended states have the average symmetry of the pure system.  In the present problem, however,  the $t_{ij}$
are frustrated as a function of separation. 
{\textcolor{black}{This allows for the largest interactive binding energy compared with the local displacement cost to be for an unconventional order, as is the case for spin glasses with good local moments in the presence of a sufficient combination of quenched disorder and frustration.}
{\textcolor{black}{Thus, for a displacive system,}}
depending upon 
the {\textcolor{black}{level of }}disorder
{\textcolor{black}{and frustration}}, the true phase transition can be  to {\textcolor{black}{(i)} a phase that is periodically ordered on average or {\textcolor{black}{(ii)}} to a spin glass-like phase  {\textcolor{black}{with $p=\overline{\langle u \rangle}=0$ but $q=\overline{{\langle u \rangle}^{2}} \neq 0$}  or {\textcolor{black}{(iii)}} there can be 
no {\textcolor{black}{cooperative}} order at all. \textcolor{black}{This is in accord with observations on} 
$\mathrm {Ba(Zr}_{1-x}\mathrm{Ti}_{x})\mathrm{O}$ \cite{Maiti, Miga} where the interaction term is spatially frustrated but the pure $x=1$ system is ferroelectric, that the ordering is found to be ferroelectric for $x \geq 0.75$, but  for $0.65 \geq x \geq 0.25$ the ordering is as a relaxor. \footnote {There is a similar transition in a conventional spin glass system as a function of the magnetic concentration between average periodic order (ferromagnetic) and (amorphous) spin glass order.}
The absence of any ordered phase for $x \leq 0.25$ suggests that, for this concentration range, the upper mobility edge is not reached as $T \to 0$ and there is no true phase transition,
as 
{\textcolor{black}{as is the situation}
in pure $\mathrm {Ba}\mathrm{Zr}\mathrm{O}_3$ (BZ).
Precursor localized displaced nanoregions  are expected {\textcolor{black}{for $0<x<1$}}.

{\textcolor{black}{Note that it is not proven here that the intermediate concentration  phase will be a pseudo-spin glass but it is strongly suggested as a reasonable possibility given (i) the similarity of Equation (\ref{eq:Hcsg}) and Equations (\ref{eq:HDR}) and (\ref{eq:SFDR}) and (ii) the fact that interactions in relaxor systems are frustrated by long-range Coulomb/dipolar interactions as well as short range interactions and (iii)  magnetic systems have provided ample evidence of the existence of spin glasses for a number of different types of frustrated interaction and quenched disorder.}}

By comparison with the case of spin glass systems, the paraelectric-to-ferroelectric transitions    are expected to be sharp, with little dependence on probe frequency or history-dependence, whereas relaxor transitions ({\it{c.f.}} spin glass) are expected to be sluggish with significant probe-frequency-dependence and history-dependence  (as observable in a separation of FC and ZFC static susceptibilities \cite{Nagata, Levstik}).

The glassiness and its consequence on history-dependence of conventional spin glasses can be studied within a more sophisticated mean field theory in terms of the recognition of the existence of many `pure  states' and the identification of the (low-field limit) FC and ZFC susceptibilities in terms of different `overlap' contributions \cite{MPV}.
{\textcolor{black}{For example, for }}
the SK $\pm1$ Ising spin glass,
\begin{equation}
\chi^{FC}=T^{-1}(1-\int
_0^1
 dx q(x));
\hspace{0.5 cm} \chi^{ZFC}=T^{-1}(1-q(x=1))
\label{susc}
\end{equation}
where
\begin{equation}
dx/dq = P(q) =\sum_{S,S'}\delta(q-q_{S,S'}); \hspace{0.5 cm} q_{S,S'}= N^{-1}\sum_{i}<\sigma_{i}>_{S}<\sigma_{i}>_{S'}
\label{x(q)}
\end{equation}
and the $S,S'$ are pure states; $q_{S,S'}$ is known as the overlap between the pure states $S$ and $S'$. The existence of a non-trivial distribution of pure states implies glassy dynamics and can be related to the dynamical breakdown of conventional equilibrium statistical mechanical connections between fluctuations and dissipation (see {\it{e.g.}} \cite{Parisi_Stealing}). {\textcolor{black}{It also leads to the concept of a hierarchy of timescales \cite{Sompolinsky} and hence a severe frequency-dependence of peaks in the susceptibility \cite{Hertz83}.}} However, the reader is referred to the extensive spin glass literature for further details.

{\textcolor{black}{Note that the higher energy localized states of the Anderson model with the parameter identification of eqn. (\ref{eq:Identification}) correspond to statistically clustered regions with higher than average Ti concentration (sites with higher $\epsilon$) and that, following BT,  internally they will have dominantly $\phi$ of the same sign, so that the corresponding regions for eqn. (\ref{eq:Min F(u)})  will have a tendency towards internal ferroelectric ordering, hence {\bf{polar}} nano-regions. }

{\textcolor{black}{Turning now to the localized states above the upper mobility edge of the Anderson analogue, it is tempting to approximate by
\begin{equation}
u_{i}=0; \hspace{0.1 cm} E<0, \hspace{0.5cm}
|u_{i}|\approx  \{E/[2\tilde{\lambda}]\}^{1/2} {\rm{on}} \hspace{0.1 cm}{\rm{sites}} \hspace{0.1 cm} {\rm{with}}\hspace{0.1 cm}{\rm{eigenstate}} \hspace{0.1 cm} \phi_{i} \neq 0; \hspace{0.1 cm} E>0,
\label{mfu}
\end{equation}
}} {\textcolor{black}{However, in fact account must be taken of the polarizations associated with condensation of higher E eigenfunctions in assessing the potential for condensation of local polarizations associated with lower E. In a pure system, where the eigenstates are extended, this normally results in suppression of the ordering associated with the lower eigenstates. In the case of localized eigenstates however there exists the possibility that there may still remain the many separated localized polarization clusters, with effective interactions between them that are too weak to cause coalescence to an extended state 
above a characteristic temperature, at which 
there can be a  transition to a phase which may be either periodic on average or relaxor, depending upon the energetic balance. Explicit solution of Equation (\ref{eq:Min F(u)}), or a simplification - see the next section -  would be valuable to assess this further within the context  of mean field theory, but the observation of PNRs in both experiment and simulation appear to demonstrate that  such localized states do exist at intermediate temperatures, while the observation of the onset of non-ergodicity at a lower temperature suggests the relaxor interpretation as a pseudo-softspin glass.}}

PNRs will contribute to deviations from pure paraelectricity, so it is tempting to identify $T_A$ with the Burns temperature $T_B$.  Caution is needed in a precise identification of the temperatures, 
{\textcolor{black}{not only because the theory discussed is only for static equilibrium and even then in mean field theory, but also}} in the absence of detailed investigation of the form of the density of states near the band edges and the weightings of localized states to the susceptibility. 
{\textcolor{black}{However, it seems probable that}}
 $T_B$ 
 measures  the temperature region where the 
{\textcolor{black}{mean field}}
localized states start to have sufficient weight to contribute significantly and the conceptualization provides a useful starting point for further theoretical modelling and further simulational investigation, beyond that in \cite{Akbarzadeh2012, Prosandeev}. It is also notable that the Burns temperature in BZT is comparable with the ferroelectric transition temperature in BT, which is likely to be close to where the upper limit  of the Anderson density of states crosses zero.

Note that this simple picture already
{\textcolor{black}{provides for}} 
the three characteristic features attributed to a displacive relaxor \cite{Kleemann2014b}; (i) significant frequency-dependence of the temperature of peaking of the susceptibility, with higher frequencies peaking at higher temperatures, (ii) absence of overall global spontaneous polarization or other structural symmetry-breaking and (iii) longish-lived polar nanoclusters beneath a higher temperature.

Although BZT has been taken as an example, the picture presented here can be applied to other isovalent relaxor systems, such as $\mathrm {Ba(Sn}_{1-x}\mathrm{Ti}_{x})\mathrm{O}_3$ (BST) \cite{Kleemann2014b}. Indeed, the observation that the critical concentration $x$ for crossover from ferroelectric to relaxor is very similar in BZT and BST is in accord with the expectation that both Zr and Sn are effectively immobile since the ionic radii of Zr 4+ (86 pm) and Sn 4+ (83 pm) are almost identical and significantly larger than (mobile) Ti 4+ (74.5 pm) \cite{Ionic radii}.

\section{Random site to random bond}

Above we have concentrated on random site-occupation. In the conventional local moment spin glass model, Equation (\ref{eq:Hcsg}),  the summation is over only the magnetic sites. {\textcolor{black}{In the case of a  displacive ferroelectric/relaxor  with one of the constituents having a $\kappa$ too large to allow order when pure, such as Zr in BZT, 
 then in the corresponding Equation (\ref{eq:HDR}) one can restrict the  summmation to the potentially ordering sites, Ti in the case of BZT.}} One can then rewrite the relevant Hamiltonians in terms of the potentially ordering ions only by relabelling the latter as $\{k,l\}$ to yield
 random bond forms, summed on all (new) sites. {\textcolor{black}{Thus for the spin glass system}}
\begin{equation}
H_{RBSG}=-\sum_{(kl)}  {J}_{kl}{\textit{\textbf{S}}}_{k}.{\textit{\textbf{S}}}_{l}
\label{eq:HRBSG}
\end{equation}
and {\textcolor{black}{for the relaxor system}}
\begin{equation}
H_{RBDR}=\sum_{k}
\{\kappa| \textit{\textbf{u}}_{k} |^2
+\lambda|\textit{\textbf{u}}_{k}|^4
\}
+ {\sum_{k>l}
J_{kl}{\textit{\textbf{u}}}_{k} {\textit{\textbf{u}}}_{l}},
\label{eq:HRBDR}
\end{equation}
where the $J_{kl}$ are now quasi-random, actually determined by the  $\{J_{i_{k}j_{l}}\}$ where $i_{k}$, $j_{l}$ are the original {\textcolor{black}{quenched-random magnetic/displacable}} site-labels of the newly labelled ${k,l}$.
In their famous innovative work on spin glasses Edwards and Anderson (EA) \cite{EA} replaced the actual $J_{kl}$ distribution by a truly random (i.i.d.) one, 
characterised by a distribution $P(J_{kl})$ with the same form for each similar site-separation (in the $\{kl\}$ lattice representation). Most subsequent theoretical and simulational work has used this random-bond formalism, usually restricted to nearest neighbours. This work has demonstrated that the essential physics is preserved.

Furthermore, to highlight the possibility of a completely new type of cooperative order EA concentrated on the symmetric case $P(J_{kl}) = P(-J_{kl})$, which excludes any average periodic order, ${\it{i.e}}$ any conventional order. Again, subsequent work has demonstrated that the spin glass behaviour is preserved and hence that the spin glass is truly a new phase compared with conventional periodic ordered states.

Similar philosophies can be used for the relaxor problem, again replacing the actual experimental quasi-random $J_{kl}$ distribution by a theoretical i.i.d. random $P(J_{kl})$, with the expectation of retaining the fundamental physics. In particular, we note that for $P(J_{kl}) = P(-J_{kl})$ one might expect a relaxor to be the only cooperatively ordered phase,  the analogue of Hertz's `Stoner glass' \cite{Hertz1979} and exhibiting the complex non-ergodicity and frequency-dependent susceptibility peaks of a spin glass, provided the bootstrapping energy is sufficient to overcome the local harmonic displacement penalty, along with precursor nano-regions quasi-frozen over longish times. 
 One could easily (and instructively) simulate such a model but this appears not to have been attempted.

\section{Soluble model}

{\textcolor{black}{Edwards and Anderson \cite{EA} also introduced several new mathematical techniques, concepts and approximations to study their spin glass model within a novel mean field theory.}}
Following their inspirational work,
{\textcolor{black}{in an attempt to probe further the veracity of EA's methodology, approximations, and conclusions,}}
 the present author 
introduced a 
soluble 
\footnote{\textcolor{black}{The initially proposed solution was not exact but further analysis has demonstrated both the solubility of the model and its solution.}}
extension \cite{SK, KS}  (the Sherrington-Kirkpatrick or SK model) whose study has exposed many challenges, subtleties and applications, (see {\em{e.g.}} \cite{MPV, Talagrand, Panchenko}). In particular it has demonstrated the existence of a novel kind of phase transition in a system without periodic order and marked by the onset of a novel type of non-ergodicity. This model was concerned with a system of Ising moments interacting through independently identically distributed pairwise exchange interactions, extending over the whole system \footnote{Sometimes referred to as infinite-ranged, or as mean-field.}. It can be and has been extended to other systems with discrete local variables.  In this section it is pointed out that a similarly soluble model can be formulated for systems in which the local variables are soft, of the type envisaged for displacive relaxors.

The Hamiltonian for the SK model is
\begin{equation}
H=-{\sum_{(ij)}} J_{ij} {\sigma}_{i} {\sigma_{j}};
\hspace{0.1 cm}
{\sigma=\pm 1}
\label{SK_Hamiltonian}
\end{equation}
where the sum is over all site-pairs $(ij)$ and the $J_{ij}$ are quenched random parameters i.i.d-distributed over all $(ij)$ according to
\begin{equation}
P(J_{ij})=[(2\pi )^{1/2}{\tilde J}]^{-1}\exp[-(J_{ij}-{\tilde J}_0)^2/2{\tilde J}^2].
\label{SK_distn}
\end{equation}
and $\tilde J_0$ and $\tilde J$ are scaled with the system size $N$ by
\begin{equation}
\tilde J_0 =J/N \hspace{0.3 cm} \tilde J =J/N^{1/2}
\label{SK_scaling}
\end{equation}
with $J_0$ and $J$ both intensive, leading to intensive transition temperatures.

The corresponding Hamiltonian for the soluble displacive relaxor problem is
\begin{equation}
H=\sum_{i} (\kappa u_{i}^2 +\lambda u_{i}^4) +
{\sum_{(ij)}} J_{ij} {u_{i}} {u_{j}}
\label{Soluble_relaxor}
\end{equation}
where the $u$ may take any real value and one can consider either positive or negative $\kappa$ but $\lambda$ is assumed positive to  prevent runaways and the $J$ are distributed as in the SK model. 

\section{PMN and PZT}

PMN ($\mathrm {Pb(Mg}_{1/3}\mathrm{Nb}_{2/3}\mathrm{)O}_3$) is more complicated than BZT for several reasons.

First, because it is heterovalently disordered, Mg and Nb in PMN carrying charges +2 and +5, different from the canonical +4 of Ti and Zr. This means that there are excess random charges compared with the $\mathrm {PbTi}\mathrm{O}_3$ (PT) template and consequently associated random fields. It is thus not clear {\it{a priori}} whether the dominant driver of the relaxor state is of the disordered and frustrated `exchange' {\textcolor{black}{(or effective random bond)}}   kind discussed above for BZT or  is the random fields or whether the two are both important, possibly in combination with corresponding random changes in the interactions.

Second, the coupling to overall strains is much stronger in PT than in BT, resulting in significant  enhanced bootstrapping and yielding significant strain accompanying the onset of ferroelectricity in PT; ​c/a =1.06 in the ferroelectric phase of PT compared with only 1.01 in ferroelectric BT.

Third, $\mathrm {PbZr}\mathrm{O}_3$ (PZ) is an anti-ferroelectric (see {\it{e.g.}} \cite{Tagantsev}), in contrast with paraelectric BZ.
Indeed, $\mathrm {Pb(Zr}_{x}\mathrm{Ti}_{(1-x)}\mathrm{)O}_3$ (PZT) at intermediate concentrations is a commercially important piezoelectric and has not (at least to date and within the knowledge of the author) shown any evidence of relaxor behaviour. However, the antiferroelectric displacements in PZ are principally associated with Pb displacements, 
presumably 
{\textcolor{black}{through interactions which have}} been ignored in the discussion {\textcolor{black}{of BZT}} above. 

Any comments on PMN here must therefore be more tentative until a more complete extension has been considered.

In \cite{SherringtonPSSB} the author proposed that PMN be modelled as a combination of a fictitious alloy PM*N*, in which the Mg  2+ and Nb 5+ are replaced by fictitious  ions M* 4+ and N* 4+ with the same local restoring coefficients ($\kappa$ and $\lambda$) and non-Coulombic pairwise interactions but carrying only charges charges +4 and their corresponding Coulomb interactions, together with further charges -2 and +1 at their sites, giving random fields and some extra random interactions. Comparisons of ionic radii,  Zr 4+ 86 pm, Ti 4+ 74.5 pm, Mg 2+  86 pm, and
Nb 5+ 78pm, suggest that the $\kappa$ values of Ti  4+ and Nb 5+ are similar to one another and should be relevant, while those of Zr 4+ and Mg 2+ are also similar to one another but in this case insufficient to lead to their spontaneous displacements \footnote{The $\kappa$ could be calculated using the first-principles techniques of \cite{Zhong} but the ionic radii are already highly suggestive of the similarities proposed above.}. 
Hence one might argue that bare PM*N* should behave similarly to PZT. 
In \cite{SherringtonPSSB} it was assumed that PM*N* would behave as BZT and hence already show relaxor effects, but, in the light of differences between BZT and PZT, this must now be re-assessed. The inclusion of strain coupling would seem a sensible start if one was interested in (fictitous) PM*N* {\it{per se}}. However, it is known from experiment that PMN retains a cubic structure as the temperature is lowered into the relaxor phase from the higher-temperature paramagnetic phase. Hence it seems reasonable to continue to consider the anticipated behaviour of PMN from a modelling that excludes global strain-coupling effects and in which the critical ingredient for relaxor behaviour lies in the $B$-site displacements.

First, we note that, on the basis of their ionic radii, the important displacing $B$-site ions are the Nb ions, while  the Mg ions can be considered immobile. Second, we assume that the main source of displacement of Pb  2+
ions lies in the extra local fields they experience due to the extra charges of Mg 2+ and Nb 5+ compared with M* 4+and N*  4+ \footnote{There will be second-order effects due to Nb displacements.}.

The energetic terms affected by the changes in ionic charges are the Coulomb interactions
\begin{equation}
H_{pert}^{B,B'}={\sum_{(ij)} Z_{i}Z_{j} e^{2}/{{\epsilon}
|[({\bf{R}}^{0}_{j} -{\bf{R}}^{0}_{i})+({\bf{u}}_{j}} - {\bf{u}}_{i})] |}
\label{eq:H5}
\end{equation}
where  the $\{{\bf{R}}^{0}_{i}\}$ are the positions of ion $\{i\}$ in the unperturbed system, the $\{Z_{i}\}$ are their charges
($Z=+2$ for Mg, $Z=+5$ for Nb, $Z=+2$ for Pb and $Z=-2$ for O) \footnote {Actually, in first principle calcuations there are found deviations from these idealized values, but this is ignored here since the aim is illustrative of concepts rather than quantitative accuracy.}, in units of the electronic charge $e$,  and $\epsilon$ is the electronic dielectric constant. Already included in PM*N* the $\{Z\}$ of M* and N* are taken as 4+ and now only the corrections need to be considered. There is a simplification in that on the $B$-sites the $\{{\bf{u}}_{i}\}$ need to be included only for Nb sites, those for Mg being effectively zero (immobile). The terms independent of $\{{\bf{u}}\}$ have no consequence for the ordering and can be ignored.

The extra linear order $B$-site displacement terms from Equation ({\ref{eq:H5}})
are
\begin{equation}
\Delta H_{RF} =- \sum_{i, {\rm{Nb}} \hspace{0.05 cm} {\rm{only}}} {{\bf{h}}}_{i}.{\bf{u}}_{i}
\end{equation}
where the ${\bf{h}}_{i}$ are  fields
\begin{equation}
{{\bf{h}}_{i}} = 2 \sum_{j} \frac{(5Z_{j}-16) e^{2}}{{\epsilon}}
\frac{{\hat{{\bf{R}}}}_{ij}^{0}
}{|{{\bf{R}}_{ij}^{0} 
|^2}},
\label{eq:PMN random field}
\end{equation}
with
\begin{equation}
{\bf{R}}_{ij}^{0}={\bf{R}}_{i}^{0}-{\bf{R}}_{j}^{0}; \hspace{0.5cm} \hat{\bf{R}}={\bf{R}}/|{\bf{R}}|,
\end{equation}
randomly depending upon the occupation of the $j$-sites by Mg 2+ or Nb 5+ions \footnote {Note that the contributions from the bare positions of the Pb and O ions cancel, by symmetry.}.

Equation ({\ref{eq:H5}}) also provides extra interaction terms quadratic in the displacements on the $B$-sites
{\textcolor{black}{
\begin{equation}
{\Delta} H_{int} =\sum_{i,j, {\rm{Nb}}
\hspace{0.05 cm} {\rm{only}}}
(\frac{9e^{2}}{\epsilon})
\frac{[ {\bf{u}}_{i} . {\bf{u}}_{j}
- 3 (
\hat
{{\bf{R}}}_{ij}^{0}.{\bf{u}}_{i})(\hat{{{\bf{R}}}}_{ij}^{0}.{\bf{u}}_{j})
]}
{|{{\bf{R}}_{ij}^{0} 
|^3}} .
\label{eq:extra_interaction}
\end{equation}
}}

In principle all the Hamiltonian terms discussed above contribute to the ordering of the $B$-sites in PMN; (i)  an effective `bare relaxor' Hamiltonian for PM*N* analagous to Equation (\ref{eq:HDR})  for BZT, (ii) a further random interaction contribution
${\Delta} H_{int}$
of Equation (\ref{eq:extra_interaction}) and (iii) a random field term $\Delta H_{RF}$ of Equation (\ref{eq:PMN random field}).

Let us now turn to the displacements of the Pb sites due to the random occupation of the $B$-sites by Mg ions of charge 2+ and Nb ions of charge 5+. Already effective charges 4+ are taken into account in PM*N*, so the issue is the effect of the extra charges --2 on Mg sites and +1 on Nb sites. These will all lead to effective extra fields on the Pb ions,
\begin{equation}
H_{pert}^{Pb}=-{\sum_{\alpha} }
{\bf{\tilde{h}}}_{\alpha}.
{\bf{\tilde{u}}}_{\alpha}.
\end{equation}
where the $\{{\alpha}\}$ label Pb sites with $\{{\bf{\tilde{u}}}_{\alpha}\}$ and
$\{{\bf{\tilde{h}}}_{\alpha}\}$
the corresponding displacement variables and effective extra fields. Taking account only of the nearest neighbour $B$-sites, the $\{{\bf{\tilde{h}}}_{\alpha}\}$ are given by
\begin{equation}
{\bf{\tilde{h}}}_{\alpha}=\sum_{j_{\alpha}}(2Q_{j_{\alpha}}e^{2}/{\epsilon})
\frac{({{\bf{{\tilde R}}}}^{0}_{\alpha}
-{\bf{R}}^{0}_{j_{\alpha}})}{|({{\bf{{\tilde R}}}}^{0}_{\alpha}
-{\bf{R}}^{0}_{j_{\alpha}})|^{3}}
\end{equation}
where the $j_{\alpha}$ are the $B$-neighbours of the $\alpha$ and the $Q_{j_{\alpha}}$ are the excess charges; $Q_{j}=(Z_{j}-5)$.

Each Pb has 8 nearest B-neighbours, along  $\langle 111 \rangle$ axes. They are distributed randomly between Mg 2+ and Nb 5+ in the ratios 1:2. When the two neighbours on the same axis are the same, either both Mg 2+ or both Nb 5+, their field contributions cancel. When the two neighbours on the same axis are different both lead to field contributions towards the Mg 2+ ion, which add constructively. Thus, at a Pb site, along any of the $\langle 111 \rangle$ axes the $B$-neighours yield  a field in one direction with probability  2/9,  in the opposite direction with probability 2/9 and no field contribution at all with probability 5/9. The same applies for all 4 $\langle 111 \rangle$ axes. The fields from these 4 axes add to provide the perturbation of the Pb position so as (at $T=0$) to minimise the total energy per Pb site between the harmonic displacement penalty and the field-driven displacement gain. This leads to a shell-like random distribution of Pb displacements around the pure cubic perovskite positions, as have been observed \cite{Vakrushev}. Further neighbours give smaller similar effects that also smooth the shell distribution. Note that this has nothing to do with the relaxor behaviour, beyond the suppression of effectiveness of coupling to global strain. At finite temperatures the usual Boltzmann weighting comes into play, with the local energetic contributions as above determining expectation.

For full comprehension PMN requires further fundamental consideration of the effects of the random fields and extra interactions, as well as possibly the inclusion of global strain coupling. but still the picture of an effective $B',B''$ core with frustration and disorder arising from spatial interaction-frustration coupled with site dilution appears likely to be a key ingredient. The effect of the random fields remains unclear, as to whether they are essential or just non-destructive, but some further discussion of field effects is given in the next section below.

{\textcolor{black}{Returning to PZT it is natural to ask further why no relaxor phase is observed, given the picture presented above. To the present author, a probable answer lies in a combination of the relatively high strength in Pb systems (compared with the local Pb harmonic distorsion strength) of both coupling to global uniform uniaxial strain, self-consistently bootstrapping significant $(c/a) \neq 1$ and ferroelectricity in PT, 
and interactions leading to antiferroelectricity in PZ, too strongly for a relaxor state to compete.  This then prompts a return to PMN with a speculation that the random fields (and possibly extra interactions) discussed above in this section combine with the pseudo-spin glass effects discussed earlier for BZT, to 
enable the relaxor to be preferred.}}

\section{Fields}

\subsection{Zero-field limit}
Susceptibilities are normally measured in the limit of small fields. To exhibit a divergence at a conventional  second order phase transition,
the applied field and response measure need to be `relevant', with the same symmetry as the order parameter characterizing the transition. This is observed in the ferroelectric transition in BT and BZT at higher Ti concentrations. By contrast, the peaks in conventional susceptibilities  marking the onset of relaxor or spin glass phases are not divergent since uniform fields and polarizations are not `relevant' parameters.
Rather, for spin glasses
the Edwards-Anderson spin glass order parameter is often taken as the  relevant observable with an applied random field as the relevant perturbation. The absence of periodicity in the ordering of a relaxor correlates with the absence of softening in fixed-wavevector probing \cite{Cowley}, without implying that there cannot be relevant mode softening of a more complexly structured normal mode.

\subsection{Finite uniform fields}

A finite applied field induces a finite magnetization/polarization, even in a paramagnet/paraelectric. Consequently a second-order para-ferro transition, that would be sharp in the zero-field limit, is wiped out in a finite field.

\subsubsection{Ising spin glass}

However, the Ising SK model system still shows a sharp  paramagnet to spin glass transition even in an applied field \cite{AT}. This (Almeida-Thouless, AT) transition is of a kind more subtle than envisaged by Ehrenfest; it represents a transition from an ergodic to a non-ergodic phase, with the onset of replica symmetry breaking (RSB) \footnote{\textcolor{black}{This nomenclature relates to a mathematical method of solution \cite{EA, Parisi80, MPV}.}}, 
a non-trivial distribution of overlaps and separation of FC and ZFC susceptibilities \cite{MPV}. The transition temperature is lowered by the applied field \footnote{For small field $h$ the reduction in this mean-field model scales as $h^{2/3}$.}. It is currently an issue of controversy as to whether this picture continues to apply  for short-range spin glasses, such as the Ising EA model. However, it does appear from simulations of a one-dimensional spin glass model with power law decaying $(R^{-\alpha})$  random exchange distribution that an AT transition still applies
for sufficiently small $\alpha$
\cite{Sharma-Young2011},
not just in the infinite-range SK limit.

\subsubsection{Vector spin glass}

In a vector SK system there are  further subtleties in a finite field. There is now a transition as the temperature is reduced from a paramagnet with a finite (induced) magnetization to a spin glass in the directions orthogonal to the field, a transverse spin glass \cite{GT}. This Gabay-Toulouse (GT) transition is again at a temperature reduced compared with the zero-field transition temperature \footnote{In this case the small $h$ reduction in spin glass transition temperature scales as $h^{2}$.}. The transition is accompanied by the onset of (strong) transverse non-ergodicity and a weaker induced longitudinal non-ergodicity. This behaviour crossses over at a lower temperature, comparable with that of the Ising AT transition, to also-strong longitudinal non-ergodicity \cite{CSG, DS_Japan}. As a consequence of this weak to strong longitudinal RSB crossover, the ZFC longitudinal susceptibility peak will appear quite rounded in a finite measuring field, peaking at a temperature reducing with field strength \cite{DS_Japan}. Qualitatively similar behaviour has been observed experimentally \cite{Nordblad} \footnote{For an earlier example of an experimental study of field-dependence see \cite{Chamberlain}.}.

There have been many studies of transitions from relaxor to highly polarized state as a function of the strength of an applied uniaxial field, the polarized state being referred to as ferroelectric. These have included experimental studies, mainly on PMN (e.g. \cite{Ye, Ljubljana}), and a simulational study of BZT \cite{Prosandeev}, but the present author is not aware that a possible Gabay-Toulouse effect has been considered  and hence it is suggested that this could usefully be included in the large picture; we note however (i) these systems do show  first order transitions that requires more than is in the simple SK model, probably including positive feedback due to global strain-coupling, and (ii) there are also PNR effects, which are absent from SK. The continuation of the non-ergodicity (procedure history-dependence)  of the relaxor phase, even in a finite field, has however been observed and recognized. It would be interesting to probe in displacive relaxor systems under a range of applied fields for the two characteristic temperature features predicted in \cite{CSG} and observed in \cite{Nordblad}.

\subsection{Random fields}

It is not possible experimentally to make true significant random fields in magnetic systems. Rather, studies have been mainly on diluted bipartite antiferromagnets in uniform fields, which map into effective ferromagnets with quasi-random fields under gauge transformation \cite{Aharony-Fishman}\footnote{Actually the randomness is correlated with the sublattice structure and is only of sign.}. Simulations are not so restricted but have concentrated on nearest neighbour systems without frustration in the absence of the fields. Most attention has been directed towards the modification of the critical behaviour of ferromagnets in the presence of random fields, but there still remain puzzles and disagreement. Random fields (and random temperatures) in scalar systems with purely ferromagnetic interactions have been shown not to permit an equilibrium spin glass solution \cite{Krzakala}.

By contrast, heterovalent displacive relaxor systems typically have significant random fields, without any necessity for a gauge-transform artifice. They thus potentially represent a very valuable experimental laboratory for fundamental physics of random field systems (as do some martensitic alloys \cite{SherringtonPSSB}).

For the vector SK model, fields which are random across all directions yield AT transitions \cite{Sharma-Young}. Since all directions are statistically equal, there is no analogue of the GT transition. This may be relevant to the effect of the random fields of  Equation (\ref{eq:PMN random field}) in the behaviour of PMN.

\subsection{Self-consistent mean-field modelling with fields}
The theoretical modelling discusssed above in subsection (\ref{Localization}) needs reconsideration in the presence of fields. The mean field equation (\ref{eq:Min F(u)}) will now include terms that are of zeroth order in $u$,
\begin{equation}
2\tilde{\kappa}_{i}(T) u_{i}
- {\sum_{j}\tilde{J}(R_{ij})u_{j}} -h_{i}= -4\tilde{\lambda}_{i}u_{i}^3  .
\label{eq:Min F(u) in fields}
\end{equation}
where $h_{i}$ is the field at site $i$. This leads to the induction of a non-zero $u$ at each relevant site and  prevents the simple determination of the $u_{i}$ by the direct comparison with the Anderson eigenfunction equation (\ref{eq:Anderson loc}) used in Equation (\ref{mfu}). This is not pursued here but the conceptual illustration of localized PNRs and extended ordered state picture of the ${h=0}$ case may be useful in considering the the solutions of Equation (\ref{eq:Min F(u) in fields}).

\section{Conclusion}

In this paper a case has been presented for considering the underlying physics of relaxor ferroelectrics
in comparison with that of spin glasses. It has been argued that displacive relaxors have some similarities with conventional (local moment) spin glasses and more especially much in common with simple pictures of itinerant spin glasses; (i) the existence beneath a characteric temperature region of a non-ergodic glassy phase in the presence of both disorder and frustration, provided that a sufficient binding energy energy is gained to overcome local displacement energetic penalties, (ii) significant frequency-dependence of susceptibility peaking near the onset of the relaxor phase and (iii) longish-lived but  localized nanoclusters of spontaneous displacement above the relaxor onset temperature but beneath a higher characteristic temperature. Comparisons with Anderson localization of electrons in disordered environments have played a useful role in building a picture. The discussion has been at a mean-field level, classical and static/equilibrium. Some aspects of expected dynamics can nevertheless be anticipated and extensions to proper dynamics envisaged.

The approach used has been simplistic and comparative/analogical rather than rigorous, complete or quantitative. The aim has been to stimulate thinking complementary to that conventional in the topic of relaxors, in the hope that the combination of perspectives will lead to a more complete and fruitful picture of these fascinating systems.

{\textcolor{black}{No claim is made, here or in \cite{Sherrington2013}, of originality in recognising similarities between relaxors and spin glasses and in utilising spin glass measures and methodology; see {\it{e.g.}} \cite{Viehland, Pirc}. The aim here has been to start from a basic minimal model  microscopics and attempt to demonstrate possible physical and conceptual explanations for the origin of  observed effects . \cite{Sherrington2013} does, however, appear to be the first suggestion of the Stoner glass concept as the basis for displacive relaxors. The relationship of PNRs to localization in the context of displacive relaxors was originally suggested in \cite{Nambu} (using the random bond description of Section 6), although the  use of the concept of localized clusters for itinerant magnetic alloys was much earlier \cite{Sherrington-Mihill}}}

\section{Acknowledgements {\textcolor{black}{and apologies}}}

The author thanks Prof. Roger Cowley for stimulating his interest in relaxors, Prof. Rasa Pirc for drawing his attention to the paper of Akbarzadeh et al on BZT. He also thanks  Profs. Wolfgang Kleemann, Laurent Bellaiche and Peter Gehring for comments, criticisms and references.

{\textcolor{black}{He also apologises to authors whose relevant work he has not cited. The aim has been to paint, broadbrush and concisely, a simple (probably oversimplified) but potentially stimulating picture of displacive relaxors as viewed from outside the mainstream of the topic.}}

\begingroup

\parindent 0pt

\def\enotesize{\normalsize}

\theendnotes

\endgroup

\label{lastpage}

\end{document}